\documentclass[graphicx, twocolumn, showkeys, aip, groupedaddress, superscriptaddress]{revtex4}

\usepackage{graphicx}% Include figure files
\usepackage{dcolumn}% Align table columns on decimal point
\usepackage{bm}% bold math
\usepackage[usenames]{color}
\usepackage{amsmath}
\usepackage{epstopdf}

\begin{document}

\title{Obtaining Self-similar Scalings in Focusing Flows}

\author{Joshua A. Dijksman}
\affiliation{Dept. of Physics \& Center for Nonlinear and Complex Systems, Duke University,  Box 90305, Durham, NC 27708-0305, USA}
\affiliation{Physical Chemistry and Soft Matter, Wageningen University, Wageningen, The Netherlands}

\author{Shomeek Mukhopadhyay}
\affiliation{Dept. of Physics \& Center for Nonlinear and Complex Systems, Duke University,  Box 90305, Durham, NC 27708-0305, USA}
\affiliation{Department of Mechanical Engineering and Materials Science, Yale University, New Haven, CT 06511, USA}

\author{Cameron Gaebler}
\affiliation{Harvey Mudd College, Department of Mathematics, 301 Platt Boulevard, Claremont, CA 91711, USA}

\author{Thomas P. Witelski}
\affiliation{Department of Mathematics, Duke University, Box 90320,
Durham, NC 27708-0320, USA}

\author{Robert P. Behringer}
\affiliation{Dept. of Physics \& Center for Nonlinear and Complex Systems, Duke University,  Box 90305, Durham, NC 27708-0305, USA}

\date{\today}

\begin{abstract} 
The surface structure of converging thin fluid films displays self-similar behavior, as was shown in the work by Diez et al [Q. Appl. Math 210, 155, 1990]. Extracting the related similarity scaling exponents from either numerical or experimental data is non-trivial. Here we provide two such methods. We apply them to experimental and numerical data on converging fluid films driven by both surface tension and gravitational forcing. In the limit of pure gravitational driving, we recover Diez' semi-analytic result, but our methods also allow us to explore the entire regime of mixed capillary and gravitational driving, up to entirely surface tension driven flows. We find scaling forms of smoothly varying exponents up to surprisingly small Bond numbers. Our experimental results are in reasonable agreement with our numerical simulations, which confirm theoretically obtained relations between the scaling exponents. 
\end{abstract}

 \maketitle

\section{Introduction}

Thin layers of fluid on a solid substrate display surprisingly rich dynamics, due to the interplay of forces at many lengthscales~\cite{simpson1982, 1985gennes, craster2009, 2009bonn}. Much progress has been achieved on the study of thin fluid film systems. They are mostly well characterized by the lubrication approximation of the Navier-Stokes equations. This elegant approximate formalism allows for tractable analysis of a wide range of fluid dynamics problems on many lengthscales, such as liquids spreading on flat surfaces~\cite{ehrhard1991}, inclined surfaces~\cite{huppert1982}, spin coating applications~\cite{schwartz2004, wu2006}, dam breaks~\cite{ancey2013} and geophysical~\cite{balmforth2000} contexts: ``thin'' here means that the height $h$ of the film is small with respect to the typical spreading lengthscale. The dynamics of the spatiotemporal evolution of the height field $h(x,y,t)$ is of a very general form, essentially a nonlinear diffusion equation. The equation is a generalization of the so-called porous medium equation~\cite{aronson1993} which is used to describe motion transport in soils. One interesting problem is that of `focusing flows', describing the advancing flow of a surrounding fluid to flood an interior dry spot. The simplest case would be of an axisymmetric fluid layer $h(r,t)$ impinging on a circular hole. Solutions for the porous medium equation were shown to have self-similar structure~\citep{diez1990}.
\par Self-similar solutions are described by the form
\begin{equation}
h(r,t) = f(t)F\left(\frac{r}{g(t)}\right).\label{eq:selfsim}
\end{equation} 
In this representation, $h$ is decomposed into $f(t)$, describing the time-dependent scaling of the height profile, an $F(\zeta)$ spatial profile function and $g(t)$ giving the scaling of the radial variable. Self-similar solutions occur in many contexts~\cite{barenblatt1996}.  When the scaling functions in (\ref{eq:selfsim}) can be obtained analytically from dimensional analysis and physical constraints as power laws with rational exponents, (\ref{eq:selfsim}) is called a first-kind similarity solution. However, this approach applied to the problem for converging flows does not uniquely determine values for the exponents. Being a second-kind similarity solution~\cite{barenblatt1996}, the scaling exponent must be obtained as part of the solution of a differential equation problem for the similarity profile $F(\zeta)$. For convergent viscous gravity currents, in which the main driving force is gravitational, the radius of the dry spot was predicted to collapse like $R_c(t)\propto (t_c-t)^\delta$ with $\delta=0.762...$ where $t_c$ is the point of collapse~\cite{diez1990, diez1992, aronson1993}.\\

At smaller length scales, \textit{surface tension} may also play a role in the dynamics of gravity currents. Much like gravity, surface tension also has a similar tendency to smooth out height variations in $h(r,t)$. The capillary length $l_d = \sqrt{\gamma/\rho g}$, the ratio between surface tension forces and gravity, set by surface tension $\gamma$, and $\rho g$ respectively, gives a typical lengthscale on which surface tension effects are significant. Under such conditions, surface tension gradients can dominate the spreading dynamics of the flow~\cite{Mukhopadhyay2009, bostwick2013,backholm2014}. The role of surface tension forces in thin films with considerable gravitational driving has received some attention~\cite{1996marino_cap}. The methods to estimate scaling exponents that we introduce here will allow us to map out collapse behavior for fluids driven by an arbitrary ratio of gravitational and capillary forces. We study converging viscous gravity currents with surface tension through experiments and numerics. Experimentally, we create an axisymmetric thin film by fast spin-up of a fluid in a confined container. Stopping the rotation subsequently yields a focusing
flow of the fluid to re-fill the central region. We obtain numerical results for comparison by solving the time dependent axisymmetric lubrication equation for the same system.

To compare analytic results to experiments or numerical results, one must extract such scaling exponents from either experimental or numerical surface structure $h(r,t)$ data. This is often a huge challenge due to the noise on $h(r,t)$ or the finite accuracy of numerical solutions. Obtaining power law exponents is generally difficult~\cite{clauset2009}. In the case of self-similar forms, the problem is compounded through the fact that multiple exponents have to be obtained simultaneously \footnote{Although the exponents are often constrained, as in the work by Diez.}. We present here two practical methods for obtaining such exponents from experimental or numerical data. We apply the methods to probe convergent fluid films which are driven by both gravity and surface tension. The presence of surface tension in the driving of the collapse is expected to cause deviations from the Diez scaling. We indeed observe such deviations by applying our methods over the entire range from purely gravity to purely surface tension dominated flows. We find good agreement between the methods, and find a reasonable match between numerically obtained data and experimental results. The scaling form of Eq.~\ref{eq:selfsim} can in fact even be used in the regime where both driving forces play a role in the fluid flow. Our methods are of general interest in contexts where scaling forms are to be expected and wherever experimental or numerical data can be cast in scaling form.

The article is arranged as follows: first, we discuss the practical background of converging fluid films driven by both gravity and surface tension. Then we discuss in Sec.~\ref{sec:exp} the experimental set-up used to obtain experimental data, including the setting up of initial conditions for the collapse. This also provides the context for the type of numerical analysis we will do. In Sec.~\ref{sec:num} we introduce the theoretical background, the relevant scaling parameters and the numerical methods and we elaborate on the numerical scheme used to obtain solutions to the thin film equation. The numerical results reveal the typical difficulty of extracting exponents with log-fitting of power laws, both in experimental and numerical data. We then describe the two methods to extract scaling exponents, using our numerical data as a guide. The effectiveness of the methods is then demonstrated, and used to explore the transition in dynamics from gravity driven, to surface tension driven flows. We also compare the numerical results with some experimental data. 

\section{Experiments}\label{sec:exp}

\begin{figure}[tbp]
\includegraphics[width=8.5cm]{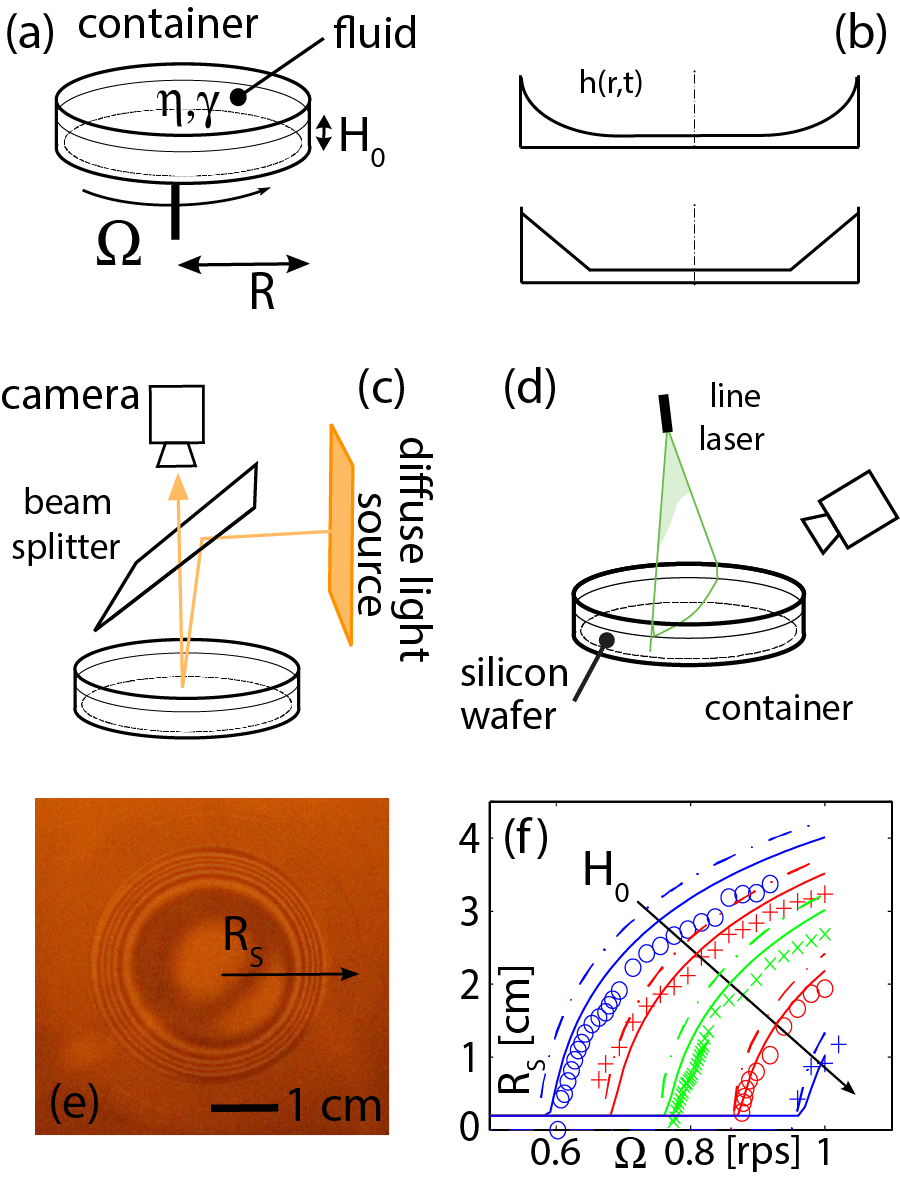}
\caption{\label{fig:setup} (Color online) (a) Schematic drawing of the container and all the relevant parameters: $\Omega$ the rotation speed, $H_0$ the initial filling height, $R$ the radius of the container, $\eta, \gamma$ the viscosity and surface tension of the fluid respectively. (b) Sketch of the initial conditions for the collapse dynamics in experiments (upper figure). These are somewhat different from the initial conditions used in the numerics (lower figure). (c) Schematic drawing of the interferometry setup. (d) Schematic of the surface structure scanning experiment. (e) A typical interferometry image from a dry spot. (f) CTF size versus rotation rate: experimental observations for $H_0 = 1.4, 1.9, 2.4, 3.2, 3.9$~mm, numerical results (solid lines, see text) and the mass conservation constraint~\cite{linden1984} (dash-dot) for comparison.}
%numerical data from Tom from May 2012, available via /software/.../critsize.m
\end{figure}

The experimental approach to study these convergent flows starts by first creating an axisymmetric reservoir of fluid in a rotating container. Fluid confinement inside the container leads trivially to the following heuristic picture: during rotation, centrifugal forces drive the fluid to the outer edge of the container. Cessation of the rotation then `releases' the fluid and creates an axisymmetric flow. There are many experimental advantages of this approach: it creates highly reproducible initial conditions in a mechanically simple and small set-up. The dependence on experimental parameters such as rotation rate, fluid viscosity and precursor film thickness can be easily tested. There is full optical access to the complete fluid film height profile, via various techniques. Determining the collapsing surface structure is critical to determining the collapse dynamics, especially the possibly self-similar structure of the fluid film close to the time of collapse.\\

The experimental system consists of an initially $\sim$1~mm thick layer of fluid in a shallow cylindrical container -- see Fig.~\ref{fig:setup}a. The container is rotated using a stepper motor with closed loop controller (Parker Zeta Drive 6104). The closed loop stepper motor controller can be programmed to run any time dependent rotation speed profile $\Omega(t)$ with a maximum of two rotations per second~(rps). The container measures 13~cm in diameter and 2~cm in height. To fix the temperature dependent viscosity $\eta$ and surface tension $\gamma$ of the fluid, the container is uniformly heated to a temperature of 24$^{\circ}$C unless otherwise noted by running water at a set temperature through the double-walled rotating axis -- for details see~\cite{Mukhopadhyay2009}. On the base of the container, a silicon wafer (University Wafers) is placed. The wafer is fixed to the base through the deposition of a small ($\lesssim 1$~ml) amount of fluid between wafer and the container. Suction force remains even after complete submersion of the wafer. The container is filled with a volume $V$ of fluid which gives an initial filling height $H_0 = V/\pi R^2$ with $R$ being the radius of the container. We use polydimethylsiloxane (PDMS) for all experiments described in this work; this fluid completely wets the silicon wafer. The transparency of the PDMS and reflectivity of the silicon wafer allows for a laser-assisted alignment of the gravity-leveled fluid surface and the silicon wafer in the container, whose orientation can be tuned by set screws. Interferometry provides access to the spatial structure in the thin film dynamics -- see Fig~\ref{fig:setup}c,e. Container illumination is provided with a uniform sodium light via a beam splitter. The spatial structure of the interference pattern of reflected and incoming light waves is recorded with a high resolution digital camera.\\

\subsection{Initial Conditions in Experiments} 

In a rotating container, a fluid will set up a parabolic surface profile to balance gravitational pressure and centrifugal forces. However, for a shallow fluid layer, above a critical rotation rate, $\Omega > \Omega_c = \sqrt{gR^2/2H_0}$~\cite{linden1984}, the solid base precludes the parabolic solution and results in a piecewise-defined height profile with a central `spot'. Outside of the spot, for $R_S<r \le R$, the approximate profile remains parabolic. Within the spot, the nature of the solution depends on the wetting properties of the container's base (the silicon wafer). For fluids that wet the base, the spot will be covered by a thin film that we call the central thin film (CTF). For non-wetting fluids, the parabolic
profiles will end at a sharp contact line at $r=R_S$.  In this paper, we discuss only the case of completely wetting fluids, which sets up a height profile as sketched in Fig.~\ref{fig:setup}b.\\
To find the initial conditions before the collapse created by the rotation, we then solve the steady state lubrication equation to obtain the complete surface profile $h(r,\Omega, H_0)$ for different rotation rates and initial volumes set by $H_0$; the CTF radii extracted from the numerics are in good agreement with the approximation from Linden~\cite{linden1984} that assumes a piece-wise continuous surface structure, consisting of only a flat layer and a parabolic surface. For such a steady state solution, mass conservation implies that 

\begin{equation}
\label{eqn:parabol}
R_S = R\left[1 - \frac{1}{R}\left(\frac{2\Omega^2H_0}{g}\right)^{1/2}\right]^{1/2}~, 
\end{equation}

\noindent in which the CTF mass and surface tension at the boundary of the container~\cite{lubarda2013} are neglected. The size of the central spot $R_S$ depends on the square root of the rotation rate~\cite{linden1984} and the total volume of fluid in the container.\\
We verify Eq.~\ref{eqn:parabol} by using interferometry to characterize the thin fluid film in the center of the container. Fig.~\ref{fig:setup}e shows a typical interferometric image of the CTF. There are several features in this image. When the container is properly leveled, the CTF is axisymmetric and flat with only modest height variations at best~\cite{birnie2010}. The fine structure of the edge of the CTF is not visible through interferometry, so we arbitrarily but consistently define the edge of the CTF by the faint ring indicated by the arrow in Fig.~\ref{fig:setup}e, which is visible in all experiments. The obvious fine structure within the ring is related to the contact line between the thin film and the parabolic solution. This fine structure is nontrivial; it is left for future work. The radius $R_S$ of the CTF for various rotation rates and fluid volumes is shown in Fig.~\ref{fig:setup}f. $R_S \sim \sqrt{\Omega}$ above $\Omega_c$ is as predicted by~\cite{linden1984}. Theoretical and numerical predictions are made with the same $R,H_0$ that produced the best fit to experimental data and produce reasonable agreement. Fig.~\ref{fig:setup}f shows also that the two theoretical methods of determining the CTF size versus $\Omega$ are consistent with each other.\\
During the process of establishing a parabolic profile, the CTF is also continuously draining fluid, making the CTF change in thickness. This drainage process is governed by a balance between centrifugal forces and viscous drag in the thinning layer. The efflux of fluid, radially outward from the CTF, slows progressively over time towards an equilibrium height profile, as described by the scaling laws first given by Emslie, Bonner and Peck (EBP)~\cite{Emslie1958}. In order to observe the collapse dynamics, it is however important to make sure that the CTF is as thin as possible. Therefore, in the numerics, we set up initial conditions as shown in Fig.~\ref{fig:setup}b, where the total volume of the triangular surface area is equal to one, which can then be rescaled with the height factor $H_0$. We choose the central layer thickness $h_{00} = 10^{-4}H_0$ sufficiently small that it does not affect the results.\\

\begin{figure}[tbp]
\includegraphics[width=8.5cm]{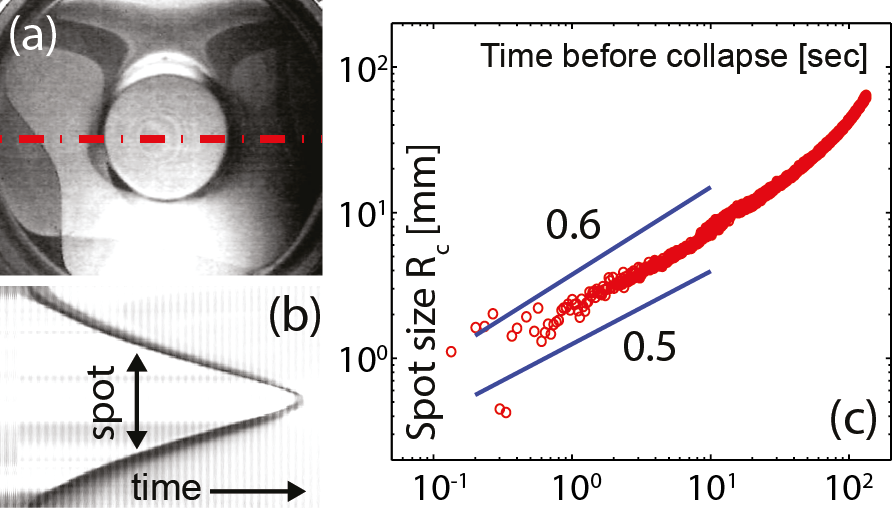}
\caption{\label{fig:topviewcollapse} (Color online) (a) Interferometric top view of a collapsing dry spot; the circular shape outlines the edge of the collapsing dry spot. The intensity profile of the dashed line is shown as a function of time in the space-time plot of panel (b), contrast enhanced. The spot size is indicated with the arrow; time progresses towards the right. Total time of collapse $\simeq$ 150 seconds. (c) The spot size as a function of the time to collapse (time increasing to the left) for a typical $H_0 = 2.9$~mm, $\eta = 10$~mPa$\cdot s$ after spinning for several minutes at $\Omega = 1$~rps. Blue lines for reference are power laws of exponents 0.5 and 0.6.}
\end{figure}

\subsection{Top View Imaging}\label{sec:topview}
\label{sec:collapse}

The fluid volume under the parabolic profile at the edge of the container serves as a reservoir that, upon cessation of the container's rotation, is `released' in a dam break fashion. This creates a fluid flux towards the center of the container. 
\par An example of this behavior is indicated in Fig.~\ref{fig:topviewcollapse}a. We use the following experimental procedure: starting from an initially flat fluid surface in a static container, the rotation rate is ramped up to $\Omega > \Omega_c$, which creates the partial parabolic profile. After rotating at constant $\Omega$ for a finite time,  rotation is stopped virtually instantaneously. This removes the centrifugal force and initiates the flow of the fluid from the reservoir at the boundary. Tracer particle tracking on the surface of the fluid showed that all experiments are at low enough Reynolds number $Re$ such that rotational flow never persists more than a small fraction of the initial collapse. The CTF spot in the center of the container thus disappears relatively slowly. We denote the size of the dynamically changing spot with $R_c(t)$ to distinguish it from the steady state spot size $R_S(\Omega)$. This process is indicated in Fig.~\ref{fig:topviewcollapse}b; these experiments are similar as the one described in Ref.~\cite{diez1992}. Note that the collapse dynamics of the front over a too-thick CTF layer ends with a standard diffusive $R_c(t) \sim (t_c-t)^{1/2}$ exponent for a fixed height threshold. This is due to the fact that with a thick CTF layer, any cavity is simply a dip in the fluid surface with an approximately parabolic shape; this dip shifts its vertical position while the fluid is moving to the center.  To avoid having this effect obscure the nontrivial scaling, it is essential to make the CTF as thin as possible just before the collapse. The container is spun sufficiently long to ensure this is the case. Imaging suggests that the typical initial film thickness is about 100 micrometer or less.\\ 

% estimate from the fact that the 3cm layer is imaged in ~30 pixels, which means 100micron per pixel. Central layer is about a pixel thick, plus the usual fuzziness. With reflection and refraction, this is always an overestimate.
%
%We measure the collapse timescale by direct imaging the CTF during collapse for a $\eta = 1000$~mPas and a range of $H = 1.7$ to $7.3$~mm. Results are shown in Fig.~\ref{fig:numsurf}b; in the accessible range of data, the experimental results are  consistent with a total collapse time scaling of $H_0^{-3}$. We discuss this scaling further in Sec.~\ref{sec:num}\\

At the end of the collapse, just before cavity closure, the spot size versus time is clearly nonlinear: from the data shown in Fig.~\ref{fig:topviewcollapse}b, the size of the shrinking CTF spot is tracked with an intensity threshold technique. The result is shown in Fig.~\ref{fig:topviewcollapse}c: the spot size varies as $R_c(t) \sim (t_c-t)^{0.55}$ with an uncertainty of about 0.05 in the exponent. However, as we will see below, a log-fitting routine is unable to capture an accurate exponent for the collapse. To better characterize the collapse dynamics, we turn to surface structure imaging.

\begin{figure}[tbp]
\includegraphics[width=8.5cm]{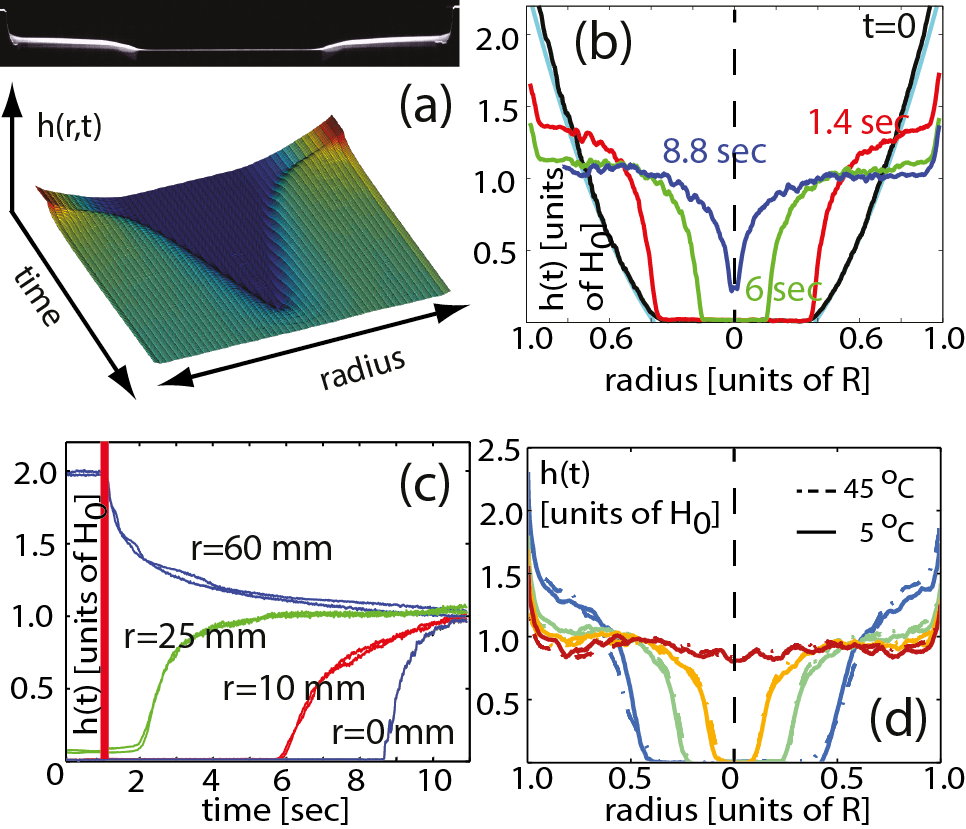}
\caption{\label{fig:collapse} (Color online) (a) Typical image (cropped) obtained by the camera. From a sequence of such images we can reconstruct the full height profile $h(r,t)$ as shown in color, indicating the height. (b) $h(r)$ for different $t$. $t=0$ refers to the rotational forcing stage which sets up the initial parabolic profile, indicated in blue, from~\citep{linden1984}. (c) $h(t)$ for different $r$ as indicated. $R_s \sim 0.38R$ for this experiment. Rotation is stopped just before $t=1$~sec. (d) As in (b) measured for the same fluid at two different temperatures. Time is rescaled with the temperature dependent viscosity ($\eta(5^{\circ}C) \sim$ 300~mPa$\cdot s; \eta(45^{\circ}C) \sim$ 36~mPa$\cdot s$) to indicate the similarity of the profiles.}
\end{figure}

\subsection{Surface Structure Imaging}

To obtain full surface structure data in the experiments, we step away from interferometry and instead image a cross section of the (radially symmetric) height profile with a high speed camera (Photron) -- see Fig.~\ref{fig:collapse}. Surface contrast is created by adding an oil-soluble fluorescent dye in the fluid (Pyrromethene 567) and illuminating the dye with a 532~nm laser line. There is sufficient dye concentration in the PDMS such that only the fluid close to the surface fluoresces (Fig.~\ref{fig:collapse}a). \\

The laser line scanning method is used to obtain the data in Fig.~\ref{fig:collapse}a-d: here a 2.9~mm layer of $\sim$ 700~mPas PDMS was spun for a minute at $\Omega = 1$~rps before cessation of rotation at $t=1.1$ seconds. The complete structure and typical behavior of the thin film surface is shown in Fig.~\ref{fig:collapse}a, much like Ref.~\cite{Bankoff2003}. The radial surface profiles, $h(r,t)$, for e.g. $t = 0, 1.4, 6$ and $8.8$ seconds are shown in Fig.~\ref{fig:collapse}b. At $t=0$ there is a parabolic surface profile due to centrifugal force, which is quantitatively in agreement with the prediction from~\citep{linden1984}. The equation describing the surface profile at $\Omega > \Omega_c$: $h(r) = \frac{\Omega^2}{g}\left(r^2 - R_S^2\right)$, with $r > R_S$. For $r < R_S$, we expect $h \sim h_{00}$, which is below the vertical resolution shown in Fig.~\ref{fig:collapse}b.

After the collapse, the parabolic surface profile first inverts its curvature to form the moving front shape as visible in Fig.~\ref{fig:collapse}b at 1.4 seconds. Called waiting-time behavior, the change in curvature happens while the contact line is essentially static~\cite{lacey1982, marino1996}. Beyond this point a symmetric collapse of the central cavity is observed. In Fig.~\ref{fig:collapse}c, $h(t)$ for $r = \pm 0, \pm 10, \pm 25$ and $\pm 60$~mm radial positions along the diameter marked by the laser line in the $65$~mm radius container is shown. Contour lines for both positive and negative $r$ are shown to indicate the radial symmetry in the surface shape; they overlap almost entirely. As shown in Fig.~\ref{fig:collapse}d, the time-dependence of the collapse dynamics scales inversely with the viscosity.\\

To extract the CTF collapse exponent, we need to measure both the size $R_c(t)$ and the collapse time $t_c$. We do this by choosing a single height threshold: we define the radius of the CTF with $$R_c(t)=\max_{h(r,t)=h_T} r,$$ the largest $r$ for which the experimentally determined height profile $h(r,t)$ is smaller than threshold $h_T$. The height threshold also sets $t_c$ with some uncertainty.  Fig.~\ref{fig:collapse}c shows that the collapse time is not easy to precisely pin-point from experimental data. The precision in determining $t_c$ is set by the slope of time dependence for the height at the origin around the collapse time. If the slope of $h(r=0, t)$ around the collapse time is large, $t_c$ does not depend sensitively on the height threshold $h_T$. We however observe in Fig.~\ref{fig:collapse}c that the slope of $h(r=0,t)$ is finite, so every $h_T$ threshold comes with its own estimate of $t_c$.\\ 

\begin{figure}[tbp]
\includegraphics[width=8.5cm]{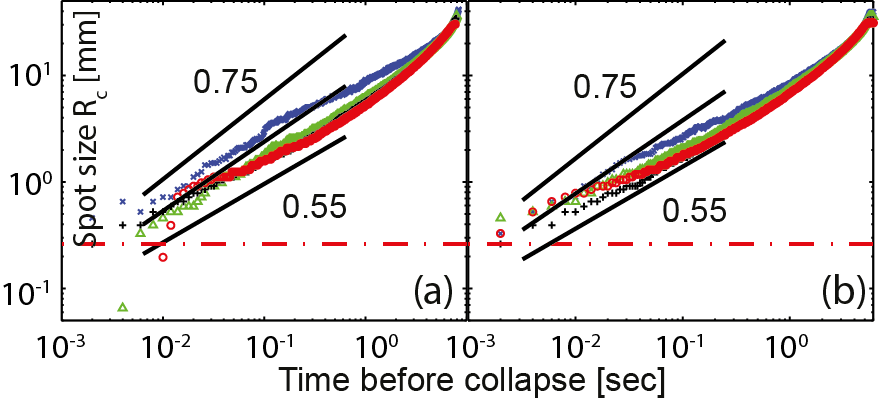}
\caption{\label{fig:expcollapse} (Color online) (a) $R_c(t)$ measurement for a $H_0=2.9$~mm PDMS layer at 24$^{\circ}C$, for which $\eta \sim$ 100~mPa$\cdot$s. For the surface profile obtained in this experiment we extract different $R_c(t)$ with different height thresholds $h_T = 0.74, 0.47, 0.37, 0.21 H_0$. (b) $R_c(t)$ measurement for a $H_0=2.9$~mm PDMS layer at 45$^{\circ}$C with $\eta \sim$ 36~mPa$\cdot$s. $R_c(t)$ obtained with different height thresholds $h_T = 0.71, 0.67, 0.57, 0.33 H_0$. In both panels,  power laws of 0.55,0.65,0.75 indicated for reference. The dash-dotted line indicates the spatial resolution of the experiment.}
\end{figure}

We show $R_c(t)$ for two different collapse experiments and four different $h_T$ for each experiment in Fig.~\ref{fig:expcollapse}a,b. For the experimental settings used, the data in Fig.~\ref{fig:expcollapse}b shows that $R_c(t) )\sim (t_c-t)^{0.65}$ with an uncertainty in the exponent of about 0.05, but very similar to the exponent obtained with the top view experiments. These two front tracking methods therefore yield consistent results. The exponents found with these two methods are however significantly different from the $0.762$ exponent expected for gravity driven fluids, suggesting that in our experiment, surface tension plays a role. To probe the role of both driving forces, we will look at the thin film equation in more detail in the next section.

\emph{Choosing $H_0, \eta$} --- As shown below, the collapse time strongly decreases with the initial layer thickness $H_0$, and as such, it is favorable to choose thicker layers to keep experimental timescales manageable. However, for thicker layers the precursor film needs prohibitively long spinning time to become sufficiently thin at the time of the collapse. One solution for this challenge is to speed up the collapse by choosing less viscous PDMS, but for fluid viscosities of less than $\sim$ 50~mPa$\cdot$s the Stokes flow approximation breaks down during the collapse~\footnote{This is evidenced by the experimental observation of a brief emergence of a protrusion of liquid at the end of the collapse.}. We are hence limited in choosing $\eta \sim 100-1000$~mPa$\cdot$s and $H_0 \sim 3$~mm. 

\section{Theory}\label{sec:num}

We consider thin film dynamics in the geometry shown in Fig.~\ref{fig:setup}a. In the low Reynolds number, creeping viscous flow limit, the time dependent film height $h(r,t)$ in the rotating container is described with a time dependent axisymmetric lubrication equation that includes surface tension, gravity, centrifugal force and disjoining pressure:

\begin{align}
\label{eq:filmfull}
\frac{\partial h}{\partial t} + \frac{1}{r} \frac{\partial}{\partial r} \left\{ \frac{\rho\Omega^2}{3\eta}r^2h^3-rh^3\frac{\partial}{\partial r} \left[ \frac{\rho g}{3\eta} h - \frac{A}{3\eta h^3} \right] \right. \nonumber\\
\qquad \left. + \frac{\gamma}{3\eta} rh^3 \frac{\partial}{\partial r} \left[
\frac{1}{r} \frac{\partial}{\partial r}\left(r
\frac{\partial h}{\partial r} \right)\right] \right\} = 0~. 
\end{align}

\noindent Here, $h(r,t)$ is the axisymmetric surface height profile depending on the radial coordinate, $r$, and time, $t$. Viscosity is given by $\eta$, $\Omega$ is the rotation rate and the surface tension coefficient is $\gamma$. The local pressure inside the fluid layer depends both on gravity and the disjoining pressure with the Hamaker constant negative for complete wetting and of order $A \sim -10^{-20}$~J --- typically disjoining pressure effects are insignificant~\cite{kim1991}. 

\par First, starting from a uniform film of height $h=H_0$, the system is
spun up with rotation rate $\Omega$. When the profile is sufficiently
close to equilibrium (and the CTF is sufficiently thin), the rotation
is stopped, so $\Omega=0$ in (\ref{eq:filmfull}) for the rest of the evolution. The concomitant simplification of Eq.~\ref{eq:filmfull} allows for nondimensionalization. A natural choice is to set $h = H_0\tilde{h}$ and $t = T\tilde{t}$. We are then left to set a length scale for the radial coordinate. We can follow Schwartz and Roy in their drop spreading analysis for spin coating context~\cite{schwartz2004} and Bertozzi \emph{et al}~\cite{bertozzi1994} and choose $r = R\tilde{r}$. This yields the following equation in the absence of rotation ($\Omega=0$), neglecting the disjoining pressure, and dropping tildes after rewriting:

\begin{align}
\label{eq:nondimfilmfull}
\frac{1}{T}\frac{\partial h}{\partial t} + \frac{\mathcal{V}}{r} \frac{\partial}{\partial r} \left\{-rh^3\frac{\partial h}{\partial r} +  \frac{rh^3}{Bo}\frac{\partial}{\partial r} \left[\frac{1}{r} \frac{\partial}{\partial r}\left(r
\frac{\partial h}{\partial r} \right)\right] \right\} = 0~, 
\end{align}

\noindent with 

\begin{equation}
\label{eq:speed}
\mathcal{V} = \frac{\rho g H_0^3}{3R^2\eta}, \qquad Bo = \frac{\rho g R^2}{\gamma}. 
\end{equation}

\noindent The alternative choice of $r = H_0\tilde{r}$ is perhaps more natural, as one would expect that the container radius is not a relevant length scale for the convergent flows in the center of the container. The $r = H_0\tilde{r}$ rescaling only modifies the prefactors:
 
\begin{equation}
\mathcal{V} = \frac{\rho g H_0}{3\eta}, \qquad Bo = \frac{\rho g H_0^2}{\gamma} .
\end{equation}

\begin{figure}[tbp]
\includegraphics[width=8.5cm]{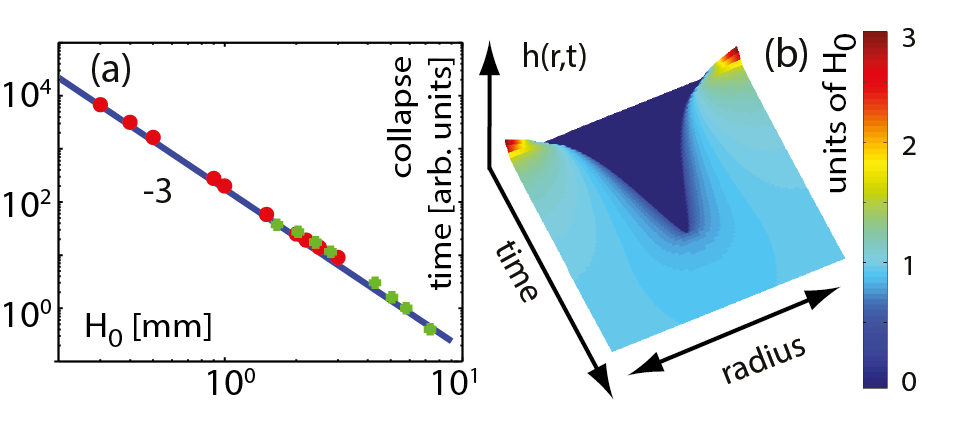}
\caption{\label{fig:numsurf} (Color online) (a) Collapse time versus filling height $H_0$; indicated is a power law with exponent $-3$. Green pluses are experimental results (see text). Red dots are numerical results. The timescale in the numerics is arbitrary; this limits us to compare only the scaling of $H_0$ with the top view experiments. (b) Typical $h(r,t)$ profile produced by the numerics. Color indicates height, in units of $H_0$. Initial conditions are as shown in Fig.~\ref{fig:setup}b.}
\end{figure}

\noindent Here the total time it takes for the collapse to complete scales with $t_c \propto H_0^{-1}$ with $t_c$ the collapse time. We compare this scaling with experiments introduced in Sec.~\ref{sec:topview}. We measure the collapse timescale by direct imaging the CTF during collapse for a $\eta = 1000$~mPas and a range of $H = 1.7$ to $7.3$~mm. Results are shown in Fig.~\ref{fig:numsurf}a; in the accessible range of data, the experimental results are consistent with a total collapse time scaling of $H_0^{-3}$, clearly incompatible with scaling from the nondimensionalization through $r = H_0\tilde{r}$.  We therefore choose $r = R\tilde{r}$, which does produce a $H_0^{-3}$ scaling of the total collapse time. The properties of PDMS, with density $\rho = 965$~kg/m$^3$ and $\gamma = 0.02$~N/m together with the container radius $R = 6.5$~cm fixes $Bo = 2\times10^3$ in all our experiments, with only a weak dependence on temperature through $d\gamma/dT \sim 0.6\times10^{-3}$~N/Km.\\

\subsection{Numerical Results}

We use a second-order-accurate implicit finite difference scheme to solve the time dependent axisymmetric lubrication equation (Eq.~\ref{eq:nondimfilmfull}). Time stepping is dynamic to allow us to resolve the fluid surface motion just before collapse. Numerics give us access to a wide range of radii $R = 0.01 \cdots 5$~m (using $\cdots$ to indicate a range) and surface tensions $\gamma = 2\times10^{-6} \cdots 2\times10^{6}$~N/m, which allow us to study a large range of $Bo = 10^{-2} \cdots 10^{10}$. The ratio of $\eta$ and the time step in the numerics is set to keep computation time small. We ensure mass conservation inside the film to within $10^{-6}H_0$ or better.\\

\begin{figure}[tbp]
\includegraphics[width=8.5cm]{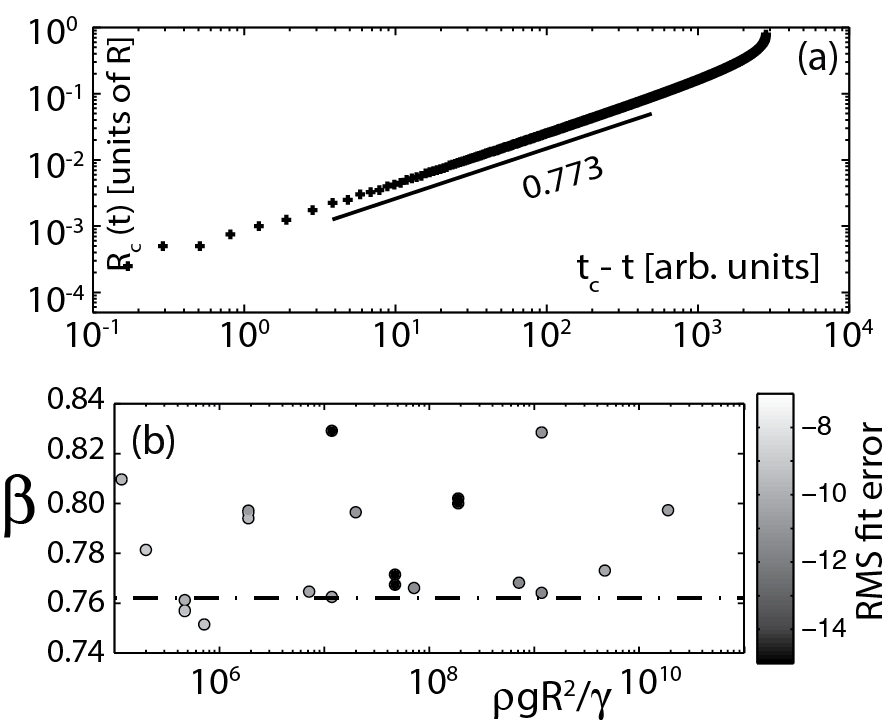}
\caption{\label{fig:numlog} (a) $R_c(t)$ for a typical numerical at $Bo = 4.7\times10^{9}$ and for threshold $h_T = 1.8\times10^{-4}H_0$. (b) Exponents from log fitting over a range of $Bo$ where the exponent is expected to be $0.762$ (dash-dotted line). The error in fitting due to $h_T$-threshold variations is of the order of the symbol size. Grayscale data points have a black border for visibility (as in other figures).}
\end{figure}

A typical $h(r,t)$ is shown in Fig.~\ref{fig:numsurf}b. The initial conditions, set up to mimic the surface profile during rotation, clearly evolve towards a collapsing CTF upon cessation of the rotation.  From these data, we extract several quantities. First, we extract the collapse time scaling by measuring the time it takes for the thin film height $h$ at $r=0.6R < R_S$ to rise to the substantial height of $0.75 H_0$, which is much larger than the initial film height $h_{00}$. The results are insensitive to the choice of these parameters, but they induce an arbitrariness in the time scale through the choice of threshold and the choice of the viscosity. We exploit this arbitrariness to shift the timescales obtained by the numerics to an order of magnitude where they overlap with experimental data. We can still extract the dependence of the total collapse time on $H_0$. The numerical collapse dynamics are indeed consistent with the experimentally observed trends shown in Fig.~\ref{fig:numsurf}a: this timescale is $\propto H_0^{-3}$, as expected from Eq.~\ref{eq:speed}.\\
Second, the size, $R_c(t)$, of the CTF or the `spot' in the center of the container is shown in Fig.~\ref{fig:numlog}a, where $R_c(t)$ is shown as a function of the time to collapse $t_c-t$ with $t_c$ the collapse time, defined as the moment at which all film heights exceed a minimum threshold level, $h(r,t_c)>h_T$.\\
The extracted scaling exponents tend to the value predicted by Diez~\textit{et. al.}~\citep{diez1992} for large values of $Bo$. However, the exponents extracted with log fitting show considerable scatter and no systematic convergence towards the $0.762$ predicted by Diez~\textit{et. al.}, indicated by the dash-dotted line in Fig.~\ref{fig:numlog}b.  We also find no trend of the extracted exponents with $H_0$ or the thresholds used for the extraction of $R_c(t)$ (not shown). This suggests that a better method to extract exponents is needed.  

\section{Extracting Exponents for Self-Similar Behavior}

To extract scaling exponents for the collapse dynamics, we can make use of the special properties of self-similar functions. Our two exponent extraction methods rely on assuming that the thin film equation Eq.~\ref{eq:filmfull} allows for similarity solutions of the form 

\begin{equation}
\label{eq:scalingform}
h(r,t)=(t_c-t)^{\alpha}F\left(\frac{r}{(t_c-t)^{\beta}}\right)~, 
\end{equation}

\noindent in which $\alpha, \beta$ are two scaling exponents, $F$ is a shape function and $t_c$ is the collapse time. In the notation of Eq.~\ref{eq:selfsim}, we thus set $f(t) = (t_c-t)^{\alpha}$ and $g(t) = (t_c-t)^{\beta}$, which means that the dry spot with diminishing size $R_c \propto g(t)$. We can use this feature to find the $\alpha, \beta$ for a particular experimental or numerical solution of the $h(r,t)$ profile, given physical variables such as surface tension, density, et cetera. In particular, we can find constraints on $\alpha(\beta)$. Plugging~(\ref{eq:scalingform}) into~(\ref{eq:nondimfilmfull}), we find that in the large $Bo$ limit, a relation 

\begin{equation}
\label{eq:alphagrav}
\alpha_{grav} = (2\beta -1)/3,
\end{equation}

\noindent is required to give self similar behavior and an ordinary differential equation (ODE) for the shape function $F$:

\begin{align}
-\frac{1}{3\zeta}\left( 3\mathcal{V} F^3F''\zeta + 9\mathcal{V} F^2F'^2\zeta + 3\mathcal{V} F^3F' - 3F'\beta\zeta^2 \right. \nonumber\\
\qquad \left.  + 2F\beta\zeta-F\zeta\right)(t_c-t)^{(2\beta-4)/3} = 0.
\end{align}

\noindent Here $\zeta = r/(t_c-t)^{\beta}$ is the reduced variable and primes denote $dF/d\zeta$, differentiation with respect to the reduced variable.  In the small $Bo$ limit, we can do the same substitution for $h(r,t)$. In this regime, we find 

\begin{equation}
\label{eq:alphacap}
\alpha_{cap} = (4\beta -1)/3,
\end{equation}

\noindent is needed to do the same separation of the shape function dynamics from the time-dependent scaling prefactor. We are thus left to find $\alpha$ and $\beta$ over the entire range of $Bo$.  

\begin{figure}[!tbp]
\includegraphics[width=8.5cm]{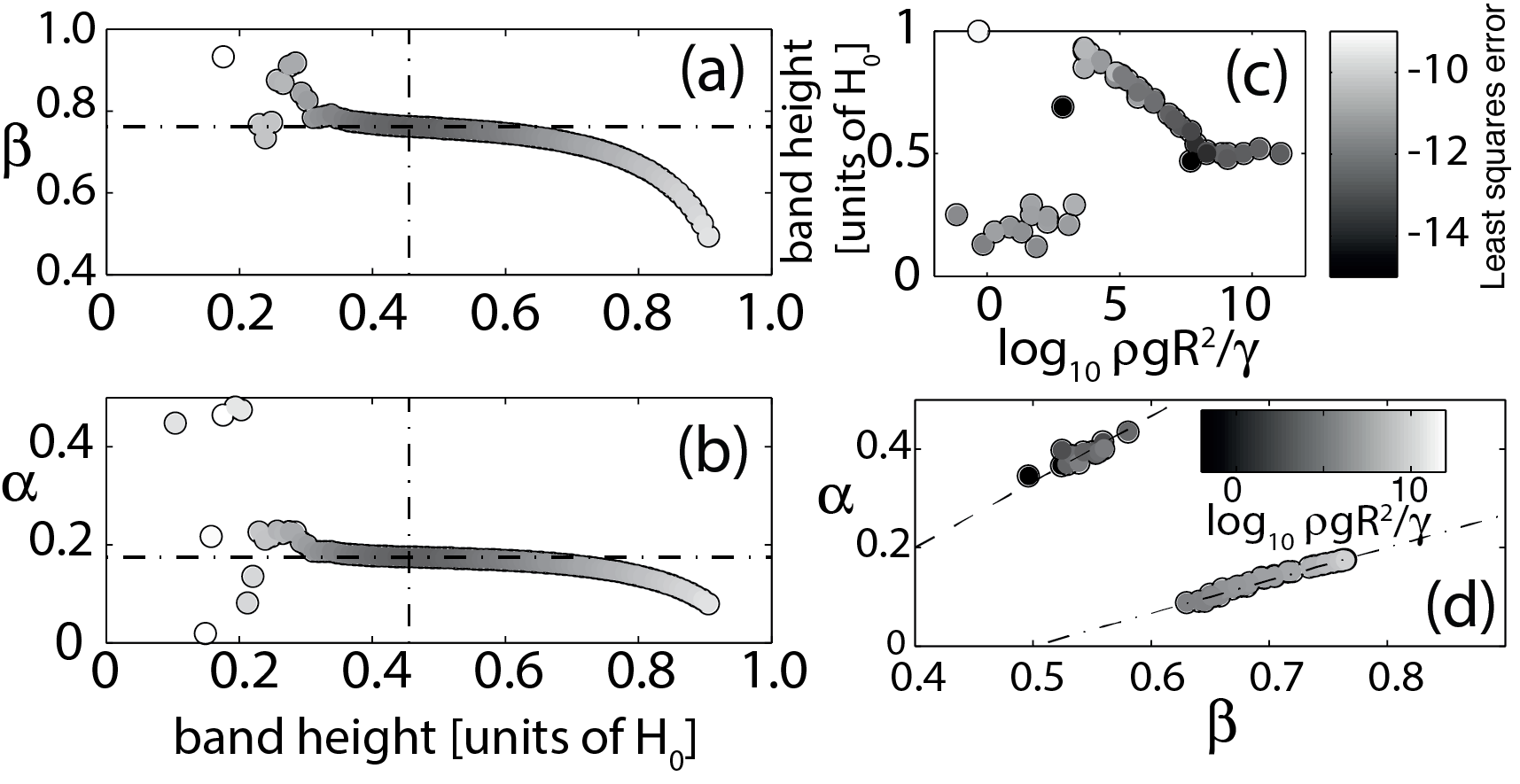}
\caption{(a,b) $\alpha, \beta$ as a function of band height, extracted with the least squares method described in the text. Grayscale indicating the $\log_{10}$ value of the nondimensionalized least squares error is shown on the right. (c) Band height for which the best $\alpha, \beta$ solution was found, as a function of $Bo$. Grayscale is the same as in (a,b). (d) the relation $\alpha(\beta)$ found with the least squares method; the dash-dotted line shows $\alpha = (2\beta-1)/3$ and the dashed line indicates $\alpha = (4\beta-1)/3$, the limiting behaviors predicted for high and low $Bo$ (indicated by gray level) respectively. \label{fig:LS_app}}
\end{figure}

\subsection{Least Squares method}

The first method to generally extract scaling exponents from numerical or experimental data relies on knowledge of the spatial and temporal derivatives $h_r, h_t$ respectively. For the solution of the scaling form of Eq.~\ref{eq:scalingform}, some algebra provides an expression relating $\alpha$ and $\beta$:

\begin{equation}
\label{eq:LSform}
(t_c-t)h_t = -\alpha h+\beta r h_r~, 
\end{equation}

\noindent This relation should be valid for all $Bo$ and at each $r,t$ if the solution is self-similar. Having access to the gradients of $h$, especially in the numerics, Eq.~\ref{eq:LSform} provides a set of linearly independent relations, with only $\alpha$ and $\beta$ as fit parameters. We can thus determine $\alpha, \beta$ to solve Eq.~\ref{eq:LSform} in a least squares sense.\\
Eq.~\ref{eq:LSform} should be valid for all $h(r,t)$ near collapse; however, we can further maximize the resolving power of this exponent extraction method by applying this relationship to \emph{height bands} within the surface. The band selection $ h_{Ti} < h < h_{Ti+1}$ selects a band of data points with a thickness of $0.01H_0$. We choose these bands anywhere between $h=0$ to $h=0.9H_0$. For each numerical data set in a particular band we compute an $\alpha, \beta$ with~(\ref{eq:LSform}). We always use the data for which $t_c-t > 0$ and limit ourselves to a time period close to the collapse, of five percent of the total collapse duration: $|t_c-t| < 0.05t_c$.  The results for a typical run are shown in Fig.~\ref{fig:LS_app}a,b for  $\beta$ and $\alpha$ respectively. In these panels, the color coding indicates the least squares error $LSE \equiv (h_t(t_c-t) + \alpha h-\beta r h_r)^2$ of all data in one height band. Both figures show similar trends: For low and high bands close to the container base and to the surface respectively, the $\alpha, \beta$ found with the least squares method give poor fits, as indicated by the large error. In an intermediate regime the error has a clear minimum. This minimum is consistently of order $10^{-14}$, as shown in Fig.~\ref{fig:LS_app}c. This panel also indicates at which band the best fit is found; the optimal band is around $0.5H_0$ for large $Bo$, moves up in the intermediate regime and is lower for small $Bo$. We verify that the least-squares method gives the relation between $\alpha$ and $\beta$ in the large and small $Bo$ limit in Fig~\ref{fig:LS_app}d, where we indicate $\alpha(\beta)$ and two limiting behavior in the dashed and dash-dotted line. Interestingly, all data is to be found on either one of the functions from Eqs.~\ref{eq:alphagrav} and~\ref{eq:alphacap} in the limiting cases, with the transition occurring around $Bo \sim 10^{4}$.\\
 
 \begin{figure}[!tbp]
\includegraphics[width=8.5cm]{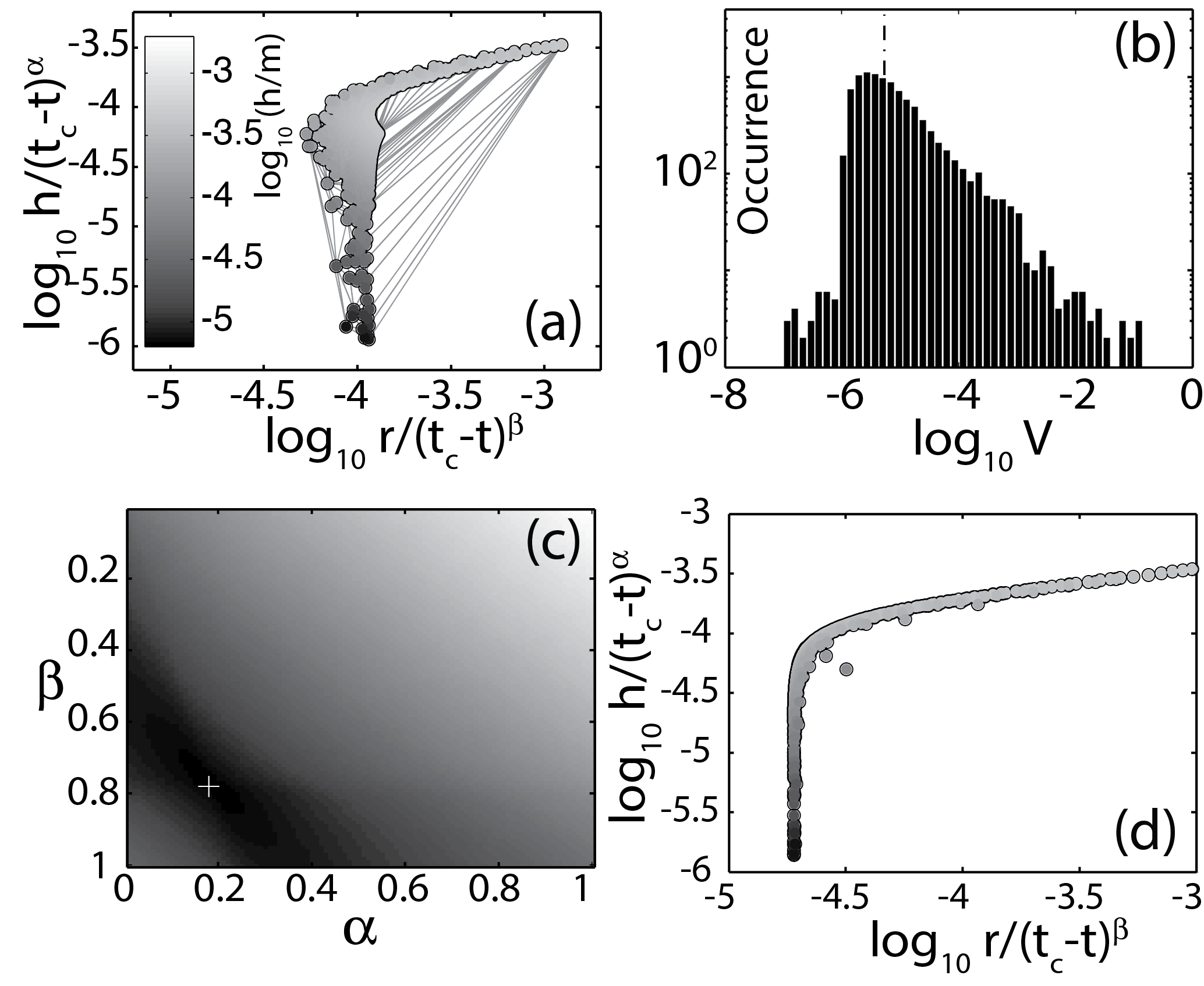}
\caption{(a) Scatter plot of rescaled $h$ obtained for $\gamma = 2\times10^{-6}$~N/m and $R= 1$~m; $Bo = 4.7\times10^9$. As an example of a combination of $\alpha, \beta$ that give a poor collapse, here we take $\alpha = 0.2, \beta = 0.6$. The light gray lines represent the triangulation of the data points. Grayscale indicates the nondimensionalized $\log_{10}(h/m)$. (b) The probability distribution function of the surface areas within the triangulation from (a). The dash-dotted line indicates the median $\tilde{V}$ of the set ${V}$. (c) $\tilde{V}(\alpha,\beta)$ for this particular numerical run. Grayscale indicates ranges from low (black) to high (white). (d) Same as (a), without triangulations and with $\alpha = 0.18, \beta = 0.78$, representing the collapse properties indicated by the $+$ in (c), the $\alpha, \beta$ for which $\tilde{V}$ is minimal. Note that we suppress dimensions in the various logarithmic axes.\label{fig:vtrimethod} }
\end{figure}

\subsection{Triangulation Method}\label{sec:tri}
The least squares method is very useful in finding scaling exponents for numerical data. However, in experimental data, obtaining derivatives $\partial_r h, \partial_t h$ is often challenging as derivatives obtained from experimental data are generally noisy. Indeed, we found that using the least squares method on experimental data does not unambiguously pick an $\alpha, \beta$ pair independently of the thresholds required in the analysis. It would therefore be advantageous to have another method to extract scaling exponents.\\
Our second method works by plotting all individual data points obtained for a measured surface profile $h(r,t)$ on a rescaled Cartesian coordinate axis $\lbrace x,y \rbrace = \lbrace r/(t_c-t)^{\beta}, h/(t_c-t)^{\alpha} \rbrace$. In general, this plotting method will produce a scattered set of data points. The crux of this method is that for the $\alpha, \beta$ that best represents the scaling function, the aforementioned scatter plot of all the surface points collapses to produce a \emph{curve} in the aforementioned coordinate system. This curve represents the function $F$ from Eq.~\ref{eq:scalingform}. For data points on this line, the Delaunay triangulation~\cite{okabe2009} of the set will produce mostly triangles with a very small area, and only a few large triangles to complete the convex hull of the entire data set. However, for a choice of $\alpha, \beta$ that does not represent the scaling behavior, the Delaunay triangulation will produce a much larger number of large triangles. By optimizing the number of small triangles in the Delaunay triangulation of the data points on the scatter plot, we can thus find the $\alpha, \beta$ pair that produces the best collapse of the data. In the Delaunay triangulation, every triangle covers an area $V$. The entire dataset makes a set of triangulated areas $\{V\}$. An example of such a set and its triangulation is shown in Fig.~\ref{fig:vtrimethod}a for \emph{numerical} results and a reasonable, but not best choice of $\alpha, \beta$. To maximize the number of small triangles, we look at the statistical properties of the set $\{V\}$. The probability distribution function for finding a triangle of area $V$ is shown in Fig.~\ref{fig:vtrimethod}b. This panel shows the distribution of areas found for the example from Fig.~\ref{fig:vtrimethod}a. We find that the median $\tilde{V}$ of the set $\{V\}$ captures the number of small triangles well: a smaller median indicates more small triangles are present in the set.  In order to find the best collapse, we thus have to compute $\tilde{V}(\alpha,\beta)$ and find its minimum.  In Fig.~\ref{fig:vtrimethod}c, we show $\tilde{V}(\alpha,\beta)$~  \footnote{The Jacobian of the rescaling coordinate transformation is generally not equal to one. This effect is taken into account in Fig.~\ref{fig:vtrimethod}c, although its effect is small.}.  The minimum of the triangulated area is indicated by the cross, whose collapse is shown in Fig.~\ref{fig:vtrimethod}d: indeed the minimum in $\tilde{V}(\alpha,\beta)$ corresponds to a very good collapse of the data. Fig.~\ref{fig:vtrimethod}d therefore also shows the form of $F$. Note that in this example, we again extract all $h(r,t)$ for which $t_c-t > 0$ and  $|t_c-t| < 0.05 t_c$, meaning all $t$ before the collapse, but sufficiently close to the collapse that the scaling assumptions should hold.\\ 

\begin{figure}[tbp]
\includegraphics[width=8.5cm]{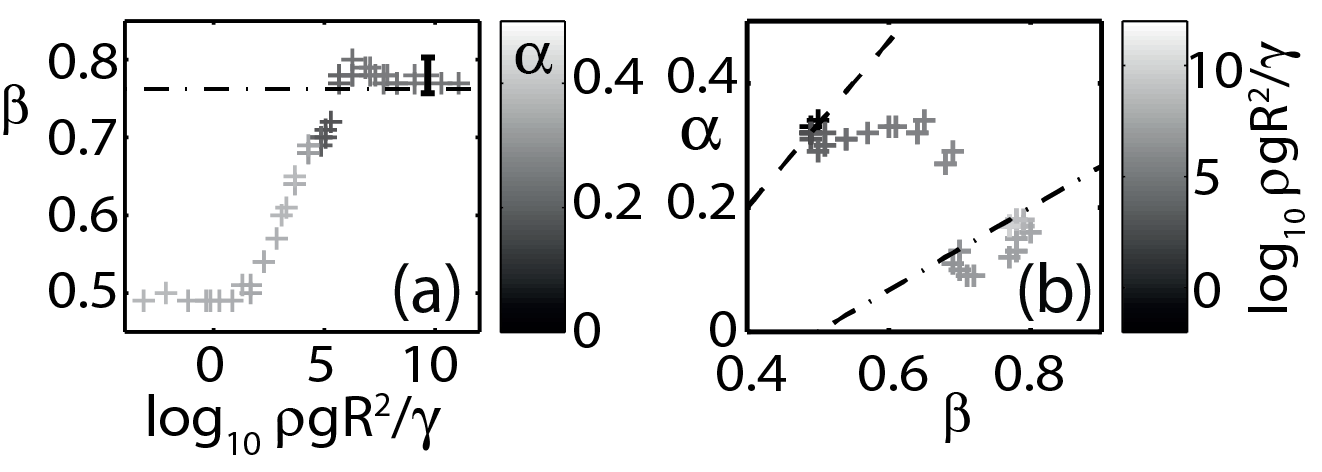}
\caption{(a) Scatter plot of $\beta$ versus $Bo$ as determined by the triangulation method. Grayscale indicates $\alpha$. The error bar indicates the uncertainty in the exponent due to the time range threshold set in the analysis. The dash-dotted line indicates the $0.762$ exponent expected in the large $Bo$ limit. (b) $\alpha(\beta)$ with data from (a); grayscale $\sim Bo$. The dashed and dash-dotted lines indicates the constraint on $\alpha(\beta)$, which are clearly satisfied in the large and small $Bo$ limit.\label{fig:vtriexponents}}
\end{figure}

To verify the triangulation method over a range of $Bo$, we apply the method to the series of numerical simulations on which we also tested the least squares methods.  We extract $\alpha, \beta$ for each of the simulations. Results are shown in Fig.~\ref{fig:vtriexponents}a,b. Fig.~\ref{fig:vtriexponents}a  shows $\beta$ versus the dimensionless ratio $\rho g R^2/\gamma$ (the grayscale indicates the value of $\alpha$). This panel shows that there are two regimes: for large $Bo$, the $\beta$ that best describe the shape of $h(r,t)$ asymptotes to the expected value of $0.762$ indicated with the dashed-dotted line. For smaller $Bo$ there is a large crossover range to a regime in which $\beta \sim 0.5$. In Fig.~\ref{fig:vtriexponents}b we show $\alpha(\beta)$. Scaling theory predicts the relationship between the exponents $\alpha(\beta)$ as indicated by the two lines. The triangulation method finds the large and small $Bo$ limiting behavior; at intermediate $Bo$, the method produces $\alpha, \beta$ that deviate from the expected constraints.\\

\begin{figure}[tbp]
\includegraphics[width=8.5cm]{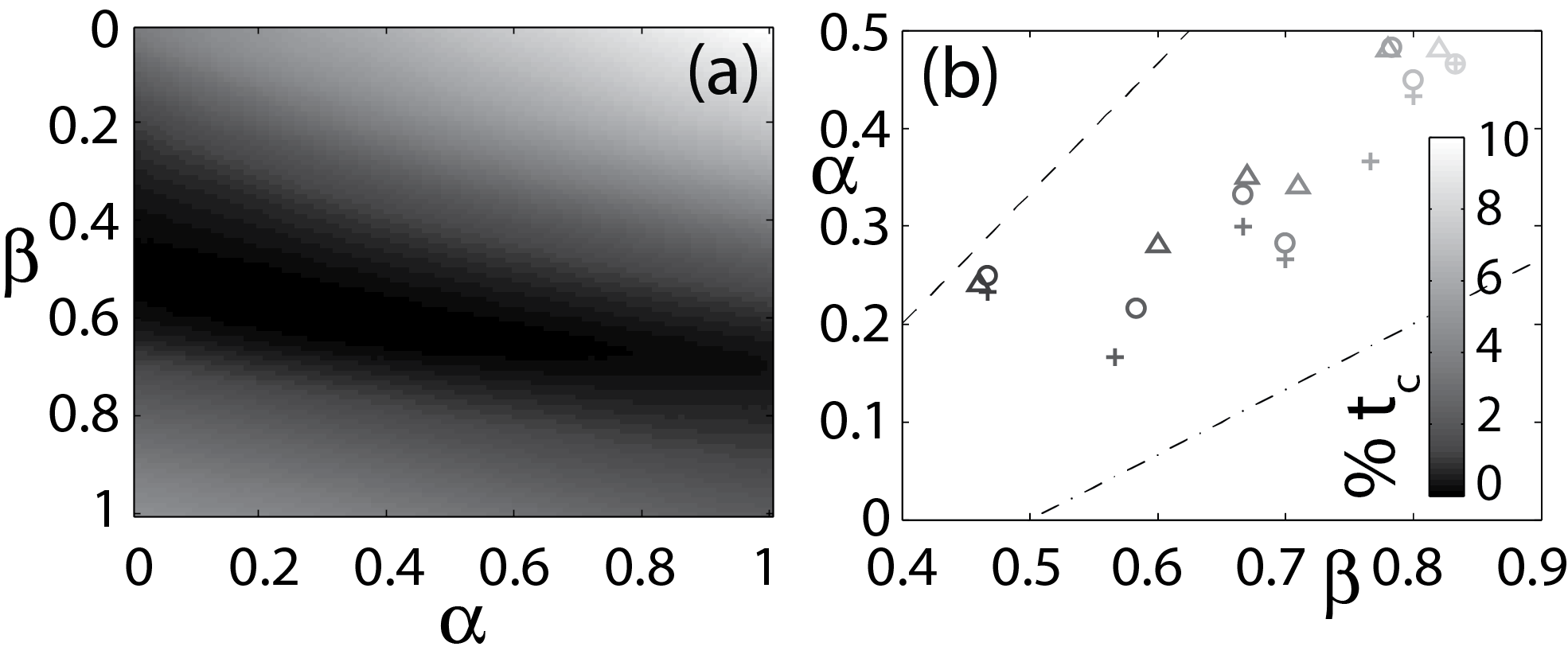}
\caption{\label{fig:Vtriexp} (a) $\tilde{V}(\alpha,\beta)$ for a particular time cutoff of 0.2 seconds for the 45~$^{\circ}C$ PDMS data set shown in Fig.~\ref{fig:expcollapse}b. The grayscale is arbitrary; black indicates a low value for $\tilde{V}(\alpha,\beta)$. (b) Minima in $\tilde{V}(\alpha,\beta)$ correspond to a pair $\alpha, \beta$. Pair location depends on $\alpha, \beta$ scan resolution: $+=0.03$, $\circ=0.16$, $\bigtriangleup=0.01$ and the percentage of the total collapse time $t_c$ included in the triangulation set (grayscale). Numerical data is always obtained with $5\%$ of $t_c$ ($0.05t_c$). Based on these thresholds, the error bars on the determination of $\alpha, \beta$ are as indicated in Fig.~\ref{fig:LS_comp}a,b.}
\end{figure}

The numerical test of the triangulation method showed that a minimum triangle area median can unambiguously be defined, which picks a unique $\alpha, \beta$ pair. We can now use the triangulation method to extract scaling exponents in experimental data.  For a fluid collapse experiment run at 45~$^{\circ}C$, the quantity $\tilde{V}(\alpha,\beta)$ for $|t_c-t| < 0.02 t_c$ is shown in Fig.~\ref{fig:Vtriexp}a. The overall structure of the experimentally determined $\tilde{V}(\alpha,\beta)$ is very similar to the numerical one; for comparison, see Fig.~\ref{fig:vtrimethod}d. The triangulation method is  sensitive to the total amount of surface structure data used in the triangulation of $F$, and for the resolution at which we test all pairs of $\alpha,\beta$ to work. This sensitivity is quantified in Fig.~\ref{fig:Vtriexp}b. We show there $\tilde{V}(\alpha,\beta)$ for several time ranges $|t_c-t| < 0.01 \cdots 0.08 t_c$ and three different resolutions at which we looked for minima in $\alpha, \beta$ space. For $0.01t_c$, the $\alpha, \beta$ pairs found approach the low $Bo$ limit function $\alpha = (4\beta-1)/3$. However, if we include more data before the collapse point into account to estimate the minimum median triangulation area, we find that $\alpha, \beta$ quickly deviate from this low $Bo$ limit. The best collapse data should be available just before the collapse, so we conservatively estimate $\alpha = 0.25 \pm 0.05$ and $\beta = 0.6 \pm 0.05$.\\

\section{Discussion}
\begin{figure}[!tbp]
\includegraphics[width=8.5cm]{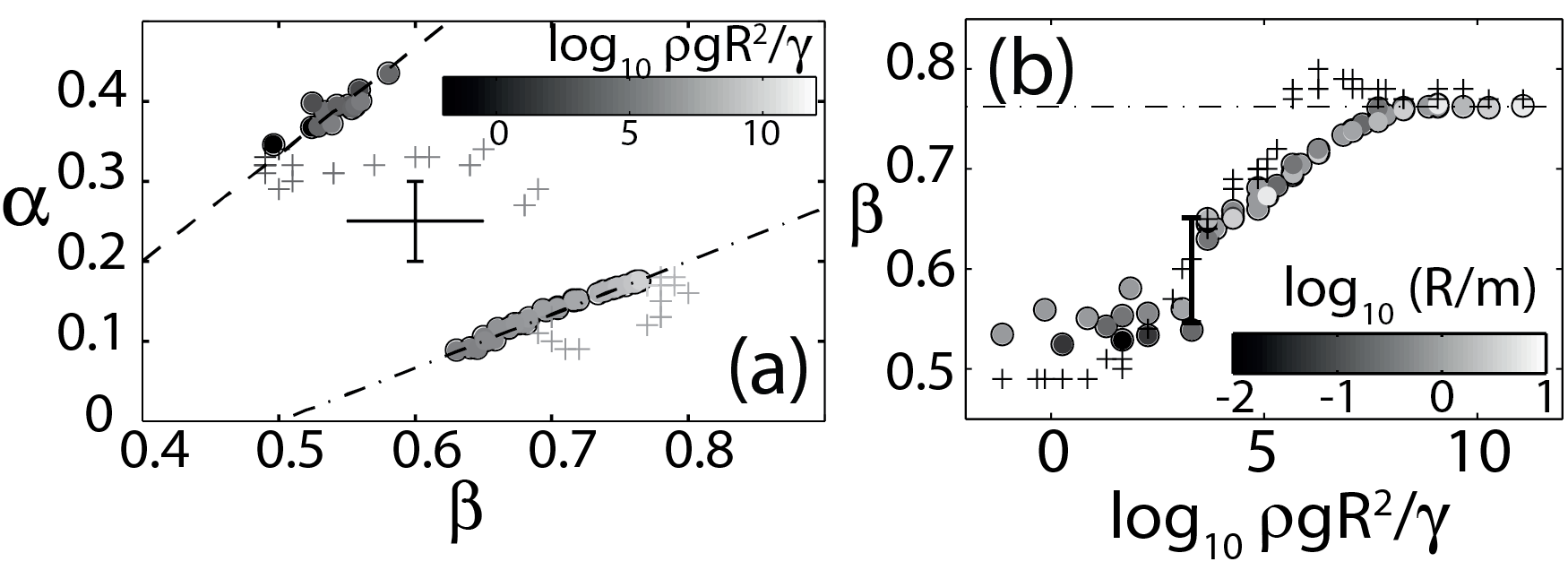}
\caption{(a) $\alpha(\beta)$ for both least squares method ($\bullet$) and the triangulation method ($+$). Grayscale indicates $\log(Bo)$; the dash-dotted line shows $\alpha = (2\beta-1)/3$ and the dashed line indicates $\alpha = (4\beta-1)/3$. The experimental data point in (a,b) comes from analysis discussed in Sec.~\ref{sec:tri}; the error bars are conservative. (b) $\beta(Bo)$ for both least squares (filled circles) and the triangulation method ($+$). The asymptotic 0.762 limit is indicated with the dash-dotted line. Grayscale indicates $\log(R)$ used in each numerical datapoint (filled circles only). The black datapoint again indicates the experimental value obtained from triangulation analysis.\label{fig:LS_comp}}
\end{figure}

We summarize the results of the analysis, the two methods to extract scaling exponents and the experimental data in Fig.~\ref{fig:LS_comp}a. There are a few important observations: \textit{(i)} the least squares method very well satisfies the derived relations for $\alpha(\beta)$ at all $Bo$, even though these relations are technically only valid in the limits of very small and very large $Bo$. \textit{(ii)} the triangulation method results are in agreement with the least squares method in the large and small Bond number limits. \textit{(iii)} the single experimental data point that we have at $Bo = 2\times 10^{3}$, although representative for many repeated experiments, only agrees with the predicted relations for the most favorable choice of $t_c$-cutoff in the analysis (see Fig.~\ref{fig:Vtriexp}b).\\
Especially point \textit{(iii)} warrants a discussion. In the triangulation method, we can only get full agreement between experiment and theoretical values expected for $\alpha, \beta$ for the most favorable cutoff in $t_c$. This can hardly be called a robust result. However, for just the exponent $\beta$, the consistency with numerical results is reasonable. Apart from the triangulation method, also the independent top-view experiments and simple thresholding techniques we described in Sec.~\ref{sec:exp} give us values for $\beta$. We obtained an exponent $0.5 < \beta < 0.7$. With $Bo = 2\times10^3$, the experiments are in an intermediate regime of Bond numbers, where $\beta$ is expected to deviate from the Diez result, as can be seen in Fig.~\ref{fig:LS_comp}b. The value obtained for $\beta$ in our experiment is thus not entirely unrealistic.\\
Nevertheless, $\alpha$ is not matching, even though the triangulation method is expected to work as per our calibration of the triangulation method on the numerical data. There are two possible explanations for the deviation: either we have missed a systematic experimental bias in our methods, or the thin film equation Eq.~\ref{eq:filmfull} does not describe the dynamics very well close to the collapse. One approximation in the thin film equation is that surface gradients are small $\partial h/\partial r \ll 1$. Close to the collapse moment, the fluid surface however forms an axisymmetric "funnel" touching the base of the container, in which this approximation seems less valid. In particular the strength of the loop curvature contribution will become relevant close to the collapse. The strength of this term in the Laplacian of the surface shape also depends on the smallness of the surface gradient terms, which in the current expression may not be fully appreciated.\\
The Bond number that we used to quantify the transition from gravity to surface tension driven dynamics does not allow one to experimentally probe a large range, as the material parameters included are difficult to vary over many orders of magnitude. We chose this Bond number definition to make the collapse timescale dependence on the filling height $H_0$ consistent with experiments. It can however be argued that just before the collapse the dynamics is indeed not anymore dependent on boundary conditions such as the total volume (and hence $R$) and that a different length scale should be chosen to nondimensionalize the thin film equations. For example, one could let the Bond number scale with $H^2$ instead of $R^2$. In these thin film applications typically $H < 0.01R$, which would push $Bo$ at least four orders of magnitude lower, and the transition point between gravity driven and surface tension driven flows to $Bo \sim 0.1$, a more reasonable number. Given the mentioned uncertainties of the curvature approximations already included in the current modeling, we leave these questions for future work.  

\section{Conclusion}
 
We studied the effect of surface tension on viscous gravity currents in a novel experimental setup, quantifying the relative strength of the surface tension with a Bond number that depends on the reservoir radius. We show that the converging surface profile is of self-similar form in the surface tension dominated regime, and we derive relations between exponents for both Bo limits. Extracting power-law exponents on log-log scale turns out to not be accurate enough in the context of the exponents of similarity solutions.  We thus develop two complementary methods to extract scaling information from the self-similar structure of the converging surface profile. These methods allow us to analyze numerical data over the complete range of purely gravity driven, to purely surface tension dominated flows, and allow for a comparison with experimental data. Our experimental data seems to suggest that the collapse regime is perhaps not well described by the thin film approximation. However, the exponent extraction methods developed work very well for the numerical data and give results in excellent agreement with the theory.

\emph{Acknowledgements} --- We thank Joshua Bostwick, Roberto Camassa, Rich McLaughlin, Omar Matar and Howard Stone for stimulating discussions, and the reviewers for their patience and valuable feedback. This project was funded by NSF DMS0968252.

%\bibliography{thintest}

\end{document}